\documentstyle[12pt]{article}
\makeatletter

\@addtoreset{equation}{section}
\makeatother

\newcommand{\beq}{\begin{equation}}
\newcommand{\eeq}{\end{equation}}
\newcommand{\beqa}{\begin{eqnarray}}
\newcommand{\eeqa}{\end{eqnarray}}

\begin{document}

\begin{center}

{\Large\bf Non-BPS Solutions of the Noncommutative $CP^1$ Model in 2+1 Dimensions}

\vspace{2cm}
{\renewcommand{\thefootnote}{\fnsymbol{footnote}}
{\large
Ko Furuta\footnote{E-mail: furuta@phys.chuo-u.ac.jp},
Takeo Inami,
Hiroaki Nakajima\footnote{E-mail: nakajima@phys.chuo-u.ac.jp}\\
and
Masayoshi Yamamoto\footnote{E-mail: yamamoto@phys.chuo-u.ac.jp}
}}

\vspace{1cm}
{\it Department of Physics,
Chuo University\\
Kasuga, Bunkyo-ku, Tokyo 112-8551, Japan}

\end{center}

\vspace{2cm}
\begin{abstract}

We find non-BPS solutions of the noncommutative $CP^1$ model in 2+1 dimensions.
These solutions correspond to soliton anti-soliton configurations.
We show that the one-soliton one-anti-soliton solution is unstable
when the distance between the soliton and the anti-soliton is small.
We also construct time-dependent solutions
and other types of solutions.

\end{abstract}

\newpage

\section{Introduction}

Noncommutative field theories naturaly arise as low-energy 
descriptions of string theory (for a review see \cite{Nek}). 
Non-perturbative dynamics of string theory was 
investigated through the study of solitons and instantons 
in noncommutative gauge theories (see e.g. \cite{Har}). 
In four-dimensional Yang-Mills theory, there exist
instantons even in the $U(1)$ case \cite{NS}. 
Non-trivial solutions are also known in noncommutative scalar theories \cite{GMS}. 
The scattering of noncommutative solitons 
was studied in Yang-Mills theory 
\cite{HIO} and in scalar theories \cite{GHS,Lind etc}. 

In lower dimensions, nonlinear sigma models
exhibit many similarities with four-dimensional 
Yang-Mills theory. In 2+1 dimensions, 
nonlinear sigma models possess soliton solutions.
The BPS solitons of the $CP^N$ model 
were extended to a noncommutative space
\cite{LLY}.
The low-energy dynamics of 
the BPS solitons in the 
noncommutative $CP^1$ model was investigated \cite{FINY}. 
In noncommutative integrable sigma models, 
time-dependent solutions 
were written down explicitly \cite{LP} 
and the scattering of solitons and
anti-solitons was studied \cite{Wo}.

In this paper, we consider the 
noncommutative $CP^1$ model in 2+1 dimensions. 
In the commutative $CP^N$ model, general static 
solutions are known \cite{DZ}.
For $N \geq 2$, there exist non-BPS solutions in addition 
to the BPS solutions. On the other hand, 
in the commutative $CP^1$ model 
no non-BPS solutions exist.
We construct non-BPS solutions of the noncommutative 
$CP^1$ model and study their dynamics. 
These solutions represent the co-existence of solitons and 
anti-solitons. 

This paper is organized as follows. 
In the following section, we summarize the properties of the 
noncommutative $CP^N$ model. 
In section 3, we construct non-BPS solutions of 
the noncommutative $CP^1$ model and investigate 
the stability of the solutions. 
We further construct
time-dependent solutions by boosting
static solutions.
In section 4, other types of solutions 
are presented. 
Finally, in section 5 we discuss future problems.

\section{The Noncommutative $CP^N$ Model in 2+1 Dimensions}

We consider the $CP^N$ model on a (2+1)-dimensional noncommutative spacetime \cite{LLY}
whose spatial coordinates obey the commutation relation
\beq
[\hat{z},\hat{\bar{z}}]=\theta>0,
\label{commutation}
\eeq
where $\hat{z}=\frac{1}{\sqrt{2}}(\hat{x}+i\hat{y})$,
$\hat{\bar{z}}=\frac{1}{\sqrt{2}}(\hat{x}-i\hat{y})$.
Since (\ref{commutation}) is the commutation relation of creation and annihilation operators,
we can identify the field with an operator in the Fock space of (\ref{commutation}).
The Lagrangian of the model is given by
\beq
L=2\pi\theta{\rm Tr}(|D_t\Phi|^2-|D_z\Phi|^2-|D_{\bar{z}}\Phi|^2),
\label{L}
\eeq
where $\Phi$ is a $(N+1)$-component complex vector
with the constraint $\Phi^\dag\Phi=1$,
and Tr denotes the trace over the Fock space.
The covariant derivative is defined by
\beq
D_a\Phi=\partial_a\Phi-i\Phi A_a,
~~~A_a=-i\Phi^\dag\partial_a\Phi,
~~~a=t,z,\bar{z},
\label{Da}
\eeq
where
\beq
\partial_z=-\theta^{-1}[\hat{\bar{z}},~],
~~~\partial_{\bar{z}}=\theta^{-1}[\hat{z},~].
\label{partial}
\eeq
The model has a global $SU(N)$ symmetry and a local $U(1)$ symmetry,
$\Phi\to\Phi g,~g\in U(1)$.

The energy of a static configuration is given by
\beqa
E&=&2\pi\theta{\rm Tr}(|D_z\Phi|^2+|D_{\bar{z}}\Phi|^2)
\nonumber\\
&\ge&2\pi|Q|,
\eeqa
where
\beq
Q=\theta{\rm Tr}(|D_z\Phi|^2-|D_{\bar{z}}\Phi|^2)
\label{Q}
\eeq
is the topological charge.
The configuration which saturates the energy bound satisfies
the BPS soliton equation
\beq
D_{\bar{z}}\Phi=0,
\label{solitoneq}
\eeq
or the BPS anti-soliton equation
\beq
D_z\Phi=0.
\label{antisolitoneq}
\eeq
The BPS (anti-)soliton solution has the positive (negative) topological charge.

In order to find solutions of (\ref{solitoneq}) and (\ref{antisolitoneq}),
it is convenient to introduce a (N+1)-component complex vector $W$
and the projection operator $P$ as
\beqa
&&\Phi=W\frac{1}{\sqrt{W^\dag W}},
\label{W}\\
&&P=W\frac{1}{W^\dag W}W^\dag.
\label{P}
\eeqa
$P$ satisfies the relations
\beq
PW=W,
~~~P^\dag=P,
~~~P^2=P.
\label{Prelation}
\eeq
In terms of $W$ and $P$ (\ref{solitoneq}) can be written as
\beq
(1-P)\partial_{\bar{z}}W=0.
\label{WPsolitoneq}
\eeq
The general solution of the BPS soliton equation (\ref{WPsolitoneq}) is given by
\beq
W=W_0(\hat{z})\Delta(\hat{z},\hat{\bar{z}}),
\label{generalsolution}
\eeq
where the components of $W_0(\hat{z})$ is polynomials of $\hat{z}$
and $\Delta(\hat{z},\hat{\bar{z}})$ is an arbitrary invertible scalar function.
The highest degree of the components of $W_0(\hat{z})$ is equal to the topological charge.
The anti-soliton solution has the same form as (\ref{generalsolution})
with the components of $W_0$ being polynomials of $\hat{\bar{z}}$.

In the $CP^N$ model some gauge transformations become singular
after the noncommutative extension.
In the commutative $CP^1$ model
$W_1={\scriptsize\left(
\begin{array}{c}
\mu z^{-1}\\
1
\end{array}
\right)}$,
where $\mu$ is a parameter,
is gauge-equivalent to the BPS soliton solution
$W_2={\scriptsize\left(
\begin{array}{c}
\mu\\
z
\end{array}
\right)}$,
but this is not true in the noncommutative case.
In the noncommutative case $\hat{z}$ is not invertible
because $\hat{z}$ has a zero eigen value.
However we can define the right inverse element
$\hat{z}^{-1}=\hat{\bar{z}}\frac{1}{\hat{\bar{z}}\hat{z}+\theta}$
satisfying $\hat{z}\hat{z}^{-1}=1$ and $\hat{z}^{-1}\hat{z}=1-|0\rangle\langle0|$.
Since
$\tilde{W}={\scriptsize\left(
\begin{array}{c}
\mu\hat{z}^{-1}\\
1
\end{array}
\right)}$
is not of the form (\ref{generalsolution}),
$\tilde{W}$ is not a BPS solution.
Moreover, $\tilde{W}$ is not a solution of the equation of motion.
In section 4 we use $\tilde{W}$ to construct a solution.

In terms of $P$ the Lagrangian is written as
\beq
L=\pi\theta{\rm{\bf Tr}}(\partial_t P\partial_t P-2\partial_{\bar{z}}P\partial_z P),
\label{PL}
\eeq
where {\bf Tr} consists of the trace over the Fock space
and that over the $(N+1)\times(N+1)$ matrix indices.
The equation of motion is
\beq
[\partial_t^2 P-2\partial_{\bar{z}}\partial_z P,P]=0.
\label{Peqmotion}
\eeq
For a static configuration this equation is written as
\beq
[[\hat{z},[\hat{\bar{z}},P]],P]=0,
\label{staticPeqmotion}
\eeq
where we have used (\ref{partial}).
The BPS equations (\ref{solitoneq}) and (\ref{antisolitoneq}) are written as
\beqa
&&(1-P)\hat{z}P=0,
\label{Psolitoneq}\\
&&(1-P)\hat{\bar{z}}P=0.
\label{Pantisolitoneq}
\eeqa
Rewriting (\ref{staticPeqmotion}) as
\beq
[\hat{\bar{z}},(1-P)\hat{z}P]+[\hat{z},P\hat{\bar{z}}(1-P)]=0,
\label{Peqmotion2}
\eeq
we see that solutions of (\ref{Psolitoneq}) satisfy (\ref{staticPeqmotion}).
Similarly solutions of (\ref{Pantisolitoneq}) satisfy (\ref{staticPeqmotion}).

\section{Non-BPS Solutions of the Noncommutative $CP^1$ Model}

In this section we consider non-BPS solutions of the noncommutative $CP^1$ model,
i.e. solutions of (\ref{staticPeqmotion})
which satisfy neither (\ref{Psolitoneq}) nor (\ref{Pantisolitoneq}).
These solutions correspond to soliton anti-soliton configurations.
We investigate the stability of the one-soliton one-anti-soliton solution.
We also construct the boosted solution.

\subsection{Soliton Anti-Soliton Solution}

We consider the $2\times 2$ projection operator of the diagonal form
\beq
P=\left(
\begin{array}{cc}
P_1&0\\
0&P_2
\end{array}
\right).
\label{Pdiagonal}
\eeq
If $P$ satisfies the equation of motion (\ref{staticPeqmotion}),
$P_1$ and $P_2$ also do.
We take $P_1$ and $P_2$ to be solutions of
(\ref{Psolitoneq}) and (\ref{Pantisolitoneq}) respectively
and hence we have
\beqa
&&(1-P_1)\hat{z}P_1=0,
\label{P1solitoneq}\\
&&(1-P_2)\hat{\bar{z}}P_2=0.
\label{P2antisolitoneq}
\eeqa
In this case $P$ satisfies the equation of motion but not the BPS equations.
In the commutative case
there exist only the trivial solutions of (\ref{P1solitoneq}) and (\ref{P2antisolitoneq}),
$P_{1,2}=0,1$,
but in the noncommutative case non-trivial solutions are known.
The solution of (\ref{P1solitoneq})
is given by \cite{GHS,LP,HLRU}
\beq
P_1=\sum_{i,j=1}^r|z^i\rangle h_{ij}^{-1}\langle z^j|,
\label{P_1}
\eeq
where
\beqa
&&|z^i\rangle=e^{\theta^{-1}(z^i\hat{\bar{z}}-\bar{z}^i\hat{z})}|0\rangle,
\label{z^i}\\
&&h^{ij}=\langle z^i|z^j\rangle,
~~~h_{ij}^{-1}h^{jk}={\delta_i}^k.
\label{h}
\eeqa
Since $1-P_2$ satisfies the BPS soliton equation (\ref{Psolitoneq}),
$P_2$ can be written as
\beq
P_2=1-\sum_{k,l=1}^s|\tilde{z}^k\rangle\tilde{h}_{kl}^{-1}\langle\tilde{z}^l|.
\label{P_2}
\eeq
Therefore $P$ takes the following form
\beq
P=\left(
\begin{array}{cc}
\sum_{i,j=1}^r|z^i\rangle h_{ij}^{-1}\langle z^j|&0\\
0&1-\sum_{k,l=1}^s|\tilde{z}^k\rangle\tilde{h}_{kl}^{-1}\langle\tilde{z}^l|
\end{array}
\right).
\label{Psolitonantisoliton}
\eeq

To see that $z^i~(i=1,\dots,r)$ and $\tilde{z}^k~(k=1,\dots,s)$
are interpreted as positions of solitons and anti-solitons respectively,
we consider the large $|z^i|$ and $|\tilde{z}^k|$ limits.
Taking the $|z^i|\to\infty$ limit and considering the finite region on the $z$-plane,
(\ref{Psolitonantisoliton}) reduces to
\beq
P=\left(
\begin{array}{cc}
0&0\\
0&1-\sum_{k,l=1}^s|\tilde{z}^k\rangle\tilde{h}_{kl}^{-1}\langle\tilde{z}^l|
\end{array}
\right).
\label{Pantisolitonlimit}
\eeq
This is the BPS anti-soliton solution.
On the other hand,
in the limit of $|\tilde{z}^k|\to\infty$,
(\ref{Psolitonantisoliton}) reduces to
\beq
P=\left(
\begin{array}{cc}
\sum_{i,j=1}^r|z^i\rangle h_{ij}^{-1}\langle z^j|&0\\
0&1
\end{array}
\right).
\label{Psolitonlimit}
\eeq
This is the BPS soliton solution.
Thus we can interpret the non-BPS solution (\ref{Psolitonantisoliton}) as the configuration
which contains $r$ solitons at $z=z^1,\dots,z^r$
and $s$ anti-solitons at $z=\tilde{z}^1,\dots,\tilde{z}^s$.
The topological charge and the energy of the solution (\ref{Psolitonantisoliton})
are the sums of the contributions of $P_1$ and $P_2$.
Since the contributions of $P_1$ and $P_2$ to the topological charge are $r$ and $-s$ respectively,
we obtain
$Q=r-s$
and
$E=2\pi(r+s)$.

\subsection{Stability}

We analyze the stability of the solution
which contains one soliton at $z$ and one anti-soliton at $z=0$
\beq
P=\left(
\begin{array}{cc}
|z\rangle\langle z|&0\\
0&1-|0\rangle\langle0|
\end{array}
\right).
\label{Ppair}
\eeq
We connect this solution to the vacuum solution
\beq
P_0=\left(
\begin{array}{cc}
0&0\\
0&1
\end{array}
\right)
\label{Pvacuum}
\eeq
by a path
\beq
P_\phi=\left(
\begin{array}{cc}
\sin^2\phi|z\rangle\langle z|&\sin\phi\cos\phi|z\rangle\langle0|\\
\sin\phi\cos\phi|0\rangle\langle z|&1-\sin^2\phi|0\rangle\langle0|
\end{array}
\right),
~~~\phi\in\left[0,\frac{\pi}{2}\right],
\label{Pphi}
\eeq
which gives $P_0$ at $\phi=0$ and $P$ at $\phi=\frac{\pi}{2}$.
The energy of the configuration (\ref{Pphi}) is
\beq
E=2\pi\theta{\rm Tr}(\partial_{\bar{z}}P_\phi\partial_z P_\phi)
=4\pi\sin^2\phi\left(1+\frac{\bar{z}z}{\theta}\cos^2\phi\right).
\label{Ephi}
\eeq
Differentiating this with respect to $\phi$ twice we obtain
\beq
\frac{\partial^2 E}{\partial\phi^2}\Bigg|_{\phi=\frac{\pi}{2}}
=8\pi\left(\frac{\bar{z}z}{\theta}-1\right).
\label{partialE}
\eeq
The energy (\ref{Ephi}) has a minimum at $\phi=0$.
When $\bar{z}z<\theta$ the energy (\ref{Ephi}) has a local maximum at $\phi=\frac{\pi}{2}$
and monotonically decreases to zero at $\phi=0$.
In this case the solution (\ref{Ppair}) is unstable
and the soliton anti-soliton pair annihilates.
When $\bar{z}z>\theta$ the energy (\ref{Ephi}) has a local minimum at $\phi=\frac{\pi}{2}$
and therefore the solution (\ref{Ppair}) is metastable in this parameter space.
We do not know whether the solution is unstable under fluctuations in other directions.

\subsection{Time-Dependent Solution}

Time-dependent solutions can be obtained by boosting static solutions $P(\hat{z},\hat{\bar{z}})$.
The solution moving in the $x(=\sqrt{2}{\rm Re}z)$-direction with the velocity $v$ is given by
\beq
P_v(\hat{z},\hat{\bar{z}},t)=P(\hat{z}',\hat{\bar{z}}'),
\label{Pv}
\eeq
which satisfies the equation of motion
\beq
[\partial_t^2 P_v-2\partial_{\bar{z}}\partial_z P_v,P_v]=0,
\label{Pveqmotion}
\eeq
where $\hat{z}'=\frac{1}{\sqrt{2}}(\hat{x}'+i\hat{y}')$,
$\hat{\bar{z}}'=\frac{1}{\sqrt{2}}(\hat{x}'-i\hat{y}')$
are given by the Lorentz transformation
\beq
\hat{t}'=\frac{t-v\hat{x}}{\sqrt{1-v^2}},
~~~\hat{x}'=\frac{\hat{x}-vt}{\sqrt{1-v^2}},
~~~\hat{y}'=\hat{y}.
\label{Lorentz}
\eeq
The spatial coordinates $\hat{z}'$, $\hat{\bar{z}}'$ obey
\beq
[\hat{z}',\hat{\bar{z}}']=\theta',
~~~\theta'=\frac{\theta}{\sqrt{1-v^2}}.
\label{Lorentzcommutation}
\eeq
The Lorentz symmetry is explicitly broken by the noncommutativity.
The moving solution is obtained by the boost accompanied by
rescaling of the noncommutative parameter \cite{BL}.

For the solution of the diagonal form (\ref{Pdiagonal}),
$P_1$ and $P_2$ can be boosted with the different velocities
$\vec{v}_1$ and $\vec{v}_2$ respectively.
We introduce the coordinates $\hat{z}_1(\hat{z}_2)$, $\hat{\bar{z}}_1(\hat{\bar{z}}_2)$
given by the Lorentz transformation with the velocity $\vec{v}_1(\vec{v}_2)$.
These coordinates obey the same commutation relation as (\ref{Lorentzcommutation})
\beq
[\hat{z}_a,\hat{\bar{z}}_a]=\theta_a,
~~~\theta_a=\frac{\theta}{\sqrt{1-v_a^2}},
~~~a=1,2.
\label{zaLorentzcommutation}
\eeq
Boosting the solution (\ref{Psolitonantisoliton}) we obtain the time-dependent solution
\beq
P_{v_1 v_2}=\left(
\begin{array}{cc}
\sum_{i,j=1}^r|z_1^i\rangle h_{1,ij}^{-1}\langle z_1^j|&0\\
0&1-\sum_{k,l=1}^s|z_2^k\rangle h_{2,kl}^{-1}\langle z_2^l|
\end{array}
\right),
\label{Pv1v2}
\eeq
where
\beqa
&&|z_a^i\rangle=e^{{\theta_a}^{-1}(z_a^i\hat{\bar{z}}_a-\bar{z}_a^i\hat{z}_a)}|0_a\rangle,
\nonumber\\
&&\hat{z}_a|0_a\rangle=0,
~~~\langle0_a|0_a\rangle=1,
\nonumber\\
&&h_a^{ij}=\langle z_a^i|z_a^j\rangle,
~~~a=1,2.
\label{za}
\eeqa
When $r=1$, $s=1$ the solution (\ref{Pv1v2}) represents a soliton and an anti-soliton moving independently
with arbitrary velocities.
In the case of $r>1$ ($s>1$) the energy density of the solution (\ref{Pv1v2}) may have multiple peaks,
but this solution is not a time-dependent multi-(anti-)soliton configuration
where (anti-)soliton peaks display relative motion
because all (anti-)soliton peaks move with a common velocity.

\section{Other Solutions}

\subsection{Other Non-BPS Solutions}

We can construct other non-BPS solutions of the form (\ref{Pdiagonal})
by taking the diagonal elements $P_1$ and $P_2$ to be solutions of the equation of motion
which do not satisfy the BPS equations.
The operator $\sum_{i=1}^k|n_i\rangle\langle n_i|~(0\le n_1<\dots<n_k)$
satisfies the equation of motion (\ref{staticPeqmotion})
but not the BPS equations (\ref{Psolitoneq}) nor (\ref{Pantisolitoneq})
except for the case of $n_i=i-1,~i=1,\dots,k$
where this operator is the solution of the BPS equation (\ref{Psolitoneq}).
For example,
choosing $P_1=|n\rangle\langle n|~(n>0)$
we have the non-BPS solution
\beq
P=\left(
\begin{array}{cc}
|n\rangle\langle n|&0\\
0&1
\end{array}
\right),
~~~n>0.
\label{Pother}
\eeq
This configuration has the topological charge $Q=1$ and the energy $E=2\pi(2n+1)$.

We can also consider the non-BPS solution
\beq
P=\left(
\begin{array}{cc}
|n\rangle\langle n|&0\\
0&|m\rangle\langle m|
\end{array}
\right),
~~~n>0,~m\ge 0.
\label{Pother2}
\eeq
For this solution we cannot find $W$ from (\ref{P}).
This is the solution of (\ref{staticPeqmotion})
but not a solution of the equation of motion derived from the Lagrangian (\ref{L}).

\subsection{Finite Size Solution}

We have constructed the solutions which have the zero size in the commutative limit.
In this subsection we consider the finite size solution which has a parameter $\mu$
related to the size of the solution.
As mentioned in section 2,
$\tilde{W}={\scriptsize\left(
\begin{array}{c}
\mu\hat{z}^{-1}\\
1
\end{array}
\right)}$
is not a solution.
The projection operator
\beq
\tilde{P}=\tilde{W}\frac{1}{\tilde{W}^\dag\tilde{W}}\tilde{W}^\dag
=\left(
\begin{array}{cc}
\hat{\bar{z}}\frac{\mu^2}{(\hat{\bar{z}}\hat{z}+\theta)(\mu^2+\hat{\bar{z}}\hat{z}+\theta)}\hat{z}
&\hat{\bar{z}}\frac{\mu}{\mu^2+\hat{\bar{z}}\hat{z}+\theta}\\
\frac{\mu}{\mu^2+\hat{\bar{z}}\hat{z}+\theta}\hat{z}
&\frac{1}{\mu^2+\hat{\bar{z}}\hat{z}+\theta}(\hat{\bar{z}}\hat{z}+\theta)
\end{array}
\right)
\label{Ptilde}
\eeq
does not satisfy (\ref{staticPeqmotion}), (\ref{Psolitoneq}) nor (\ref{Pantisolitoneq}).
We show that we can construct the solution by adding the correction to $\tilde{P}$.
We consider the projection operator
\beq
P=\tilde{P}
+\frac{1}{\mu^2+\theta}\left(
\begin{array}{cc}
\theta|1\rangle\langle1|
&-\mu\sqrt{\theta}|1\rangle\langle0|\\
-\mu\sqrt{\theta}|0\rangle\langle1|
&\mu^2|0\rangle\langle0|
\end{array}
\right).
\label{Pnonbps}
\eeq
After a little algebra we obtain
\beq
(1-P)\hat{z}P
=\left(
\begin{array}{cc}
\sqrt{\theta}|0\rangle\langle1|&0\\
0&0
\end{array}
\right).
\label{Pnonbpssolitoneq}
\eeq
This implies that $P$ is not a BPS soliton solution
but satisfies the equation of motion.
Calculating $W$ which generates $P$,
we obtain
\beq
W=\left(
\begin{array}{c}
\mu\hat{z}^{-1}\\
1
\end{array}
\right)\hat{z}
+c\left(
\begin{array}{c}
1\\
-\mu\hat{\bar{z}}^{-1}
\end{array}
\right)
|1\rangle\langle0|
=\left(
\begin{array}{c}
\mu(1-|0\rangle\langle0|)+c|1\rangle\langle0|\\
\hat{z}-c\frac{\mu}{\sqrt{\theta}}|0\rangle\langle0|
\end{array}
\right),
\label{Wnonbps}
\eeq
where $c$ is a constant.
The topological charge is
\beqa
Q&=&\theta{\rm Tr}\left[\frac{1}{W^\dag W}
((\partial_z W)^\dag(1-P)\partial_z W-(\partial_{\bar{z}}W)^\dag(1-P)\partial_{\bar{z}}W)\right]
\nonumber\\
&=&\theta{\rm Tr}
\left[\frac{\mu^2}{\mu^2+\hat{\bar{z}}\hat{z}}\left(\frac{1}{\theta}|0\rangle\langle0|
+\frac{1}{\mu^2+\hat{\bar{z}}\hat{z}+\theta}
-\left(\frac{1}{\theta}|1\rangle\langle1|
+\frac{1}{\mu^2+\theta}|0\rangle\langle0|\right)\right)\right]
\nonumber\\
&=&\theta\left[\frac{1}{\theta}
+\sum_{n=0}^{\infty}\frac{\mu^2}{(\mu^2+\theta n)(\mu^2+\theta n+\theta)}
-\left(\frac{\mu^2}{\theta(\mu^2+\theta)}+\frac{1}{\mu^2+\theta}\right)\right]
\nonumber\\
&=&1+1-1
=1,
\label{Qnonbps}
\eeqa
where we have set $c=\sqrt{\frac{\mu^2\theta}{\mu^2+\theta}}$ for simplicity
because $Q$ is independent of $c$.
The energy of this configuration is $E=2\pi(2+1)=6\pi$.

The solution (\ref{Pnonbps}) has the parameter $\mu$ of the dimension of length.
To see that this parameter is related to the size of the solution,
we consider the small $\mu$ and large $\mu$ limits.
In the limit of $\mu\to 0$, (\ref{Pnonbps}) reduces to
\beq
P=\left(
\begin{array}{cc}
|1\rangle\langle 1|&0\\
0&1
\end{array}
\right).
\label{Psmall}
\eeq
This corresponds to (\ref{Pother}) with $n=1$.
On the other hand,
in the limit of $\mu\to\infty$, (\ref{Pnonbps}) reduces to
\beq
P=\left(
\begin{array}{cc}
1-|0\rangle\langle 0|&0\\
0&|0\rangle\langle 0|
\end{array}
\right).
\label{Plarge}
\eeq
This corresponds to the solution considered in section 3
and represents a soliton anti-soliton pair sitting at the origin.
We can interpret the non-BPS solution (\ref{Pnonbps}) as the configuration
which contains a soliton of the size $\mu$
and a small soliton anti-soliton pair.
In the large $\mu$ limit the soliton spreads over the space and disappears,
and hence only the soliton anti-soliton pair exists.

\section{Discussion}

In this paper, we have constructed non-BPS
solutions of the noncommutative $CP^1$ model. 
These solutions 
can be extended to the case of $CP^N,~N\ge 2$
by embedding the solutions in the $(N+1)\times(N+1)$ projection operator of the block diagonal form.
In the commutative $CP^N$ model, all the static non-BPS solutions 
were generated from the BPS solutions \cite{DZ}. 
It is a future problem
whether we can construct 
all static solutions of the noncommutative 
$CP^N$ model.
In noncommutative gauge theories, 
(non-)BPS solutions were constructed by using
the solution generating transformation \cite{Har}.
It might be possible to find such a trasformation in the noncommutative $CP^N$ model.

We have a comment on the relation between the noncommutative $CP^N$ model
and a scalar theory on a noncommutative space (the GMS model) \cite{GMS}.
Our non-BPS solutions have been constructed by using
solitons in the GMS model. 
If one considers the $(N+1)\times(N+1)$ hermitian matrix 
instead of the scalar field in the GMS model, one gets the noncommutative $CP^N$ model in the 
low-energy limit. 

Some non-Bogomol'nyi solutions of the Yang-Mills-Higgs equations were obtained by using the non-BPS 
solutions of the $CP^N$ model \cite{Ioa}. 
It is interesting to see whether our non-BPS 
solutions can be used to construct new non-Bogomol'nyi
solutions of the Yang-Mills-Higgs system on a noncommutative 
space \cite{GroNek}.

\section*{Acknowledgements}

This work is supported partially by the Grants in Aid of Ministry of Education,
Culture and Science
(\#14540265).
H. N. is supported by a Research Assistantship of Chuo University.

\end{document}